\documentclass[aps,prb,twocolumn,superscriptaddress,amsmath,amssymb,showpacs,nofootinbib]{revtex4}

\usepackage{graphicx}% Include figure files
\usepackage{dcolumn}% Align table columns on decimal point
\usepackage{bm}% bold math
\usepackage{color}
\newcommand{\nn}{\nonumber}
\newcommand{\beq}{\begin{eqnarray}}
\newcommand{\eeq}{\end{eqnarray}}

\makeatletter
\AtBeginDocument{\@ifpackageloaded{natbib}{\ifNAT@numbers\if@filesw\immediate\write\@auxout{\string\global\string\NAT@numberstrue}\fi\fi}{}}
\makeatother
\begin{document}

\title{
Dirac-Fermion-Induced Parity Mixing in Superconducting Topological Insulators
}

\author{Takeshi Mizushima}
\affiliation{Department of Physics, Okayama University, Okayama 700-8530, Japan}
\affiliation{Department of Materials Engineering Science, Osaka University, Toyonaka, Osaka 560-8531, Japan}
\author{Ai Yamakage}
\author{Masatoshi Sato}
\author{Yukio Tanaka}
\affiliation{Department of Applied Physics, Nagoya University,
Nagoya 464-8603, Japan}

\date{\today}

\begin{abstract}
We self-consistently study surface states of 
 superconducting topological insulators.
We clarify that, if a topologically trivial bulk $s$-wave pairing
 symmetry is
 realized, parity mixing of
pair potential near the surface is anomalously enhanced  by surface
 Dirac fermions, opening an additional surface gap
 larger than the bulk one.
In contrast to classical $s$-wave superconductors,
the resulting surface density of state hosts an
extra coherent peak at the induced gap besides a
 conventional peak at the bulk gap.
 We also find that no such extra peak appears for odd-parity superconductors with a cylindrical Fermi surface.
Our calculation suggests that
the simple U-shaped scanning tunneling microscope spectrum in Cu$_x$Bi$_2$Se$_3$ does not
originate from $s$-wave superconductivity, but can be explained by odd-parity superconductivity with a cylindrical Fermi surface.
\end{abstract}

%74.45.+c Proximity effects; Andreev reflection; SN and SNS junctions
%74.20.Rp Theories and models of superconducting state: Pairing symmetries (other than s-wave)
%73.20.At Surface states, band structure, electron density of states
%03.65.Vf  Topological phases(quantum mechanics)
%74.25.Jb Properties of superconductors: Electronic structure (photoemission, etc.)

\maketitle

%-------------------- Introduction
\section{Introduction}

Topological superconductors (TSCs) are a new state of matter~\cite{Schnyder08,qi11,tanaka12,alicea12} with
nonzero topological numbers of bulk wave functions.
The topologically protected gapless surface Andreev bound states
(SABSs) are their own
antiparticles, realizing Majorana fermions in condensed matter
systems.~\cite{Wil_NP09}
Recently, the newly discovered superconductor Cu-doped Bi$_2$Se$_3$ has been considered as one potential TSC candidate.~\cite{fu10} %because of its peculiar band structure and strong spin-orbit coupling~\cite{fu10}.
The parent material, Bi$_2$Se$_3$,
is a topological insulator
with gapless surface Dirac fermions~\cite{ando2013,hasan10,qi11,hasan11}, 
but superconductivity appears by intercalating Cu.
The superconducting Cu$_x$Bi$_2$Se$_3$ retains the surface Dirac fermions
in the normal state, and thus it is dubbed
superconducting topological insulator (STI).
From the Fermi surface structure of the material,
Cu$_x$Bi$_2$Se$_3$ is predicted to be a TSC~\cite{fu10} if
time-reversal-invariant odd-parity superconductivity is
realized.~\cite{Sato09,Sato10}

Since TSCs predict gapless SABSs,
tunneling spectroscopy via SABS~\cite{TK95,kashiwaya00,kashiwayaPRL2011} can directly
access the topological superconductivity.
For Cu$_x$Bi$_2$Se$_3$, a point-contact experiment~\cite{sasaki11} has revealed a pronounced
zero-bias conductance peak (ZBCP) supporting a topological odd-parity
superconductivity.
The surface structure and tunneling spectroscopy
have been studied theoretically,~\cite{hao11,hsieh12,yamakage12,Takami} and
the ZBCP of the experiment has been reproduced theoretically.~\cite{yamakage12}
There are also several other theoretical studies about this material.~\cite{yamakage13,Yip2013,Hashimoto2013,hashimoto14,Nagai,Zocher,Chen,black-schaffer,nagai2013,brydon2014,YuanPRB2012,bitan1,bitan2,bitan3}
%Several other theoretical studies
%of this material
%\cite{yamakage13,Yip2013,Hashimoto2013,Nagai,Zocher,Chen} have also been completed. \par

Although similar ZBCPs have been observed independently,~\cite{koren11,kirzhner12,koren12}
there also have been conflicting reports of tunneling spectroscopy
recently.~\cite{peng13,Levy}
In particular, a scanning tunneling microscope (STM)
experiment has indicated a simple U-shaped
tunneling conductance for Cu$_x$Bi$_2$Se$_3$, which led to 
contrary statement that this material has
a nontopological
$s$-wave pairing symmetry.~\cite{Levy}

In this paper, we shall revisit surface states of STIs, motivated by the forementioned progress of experiments. Employing a self-consistent calculation, we shall clarify that the puzzling issue is understandable with the context of topological odd-parity pairing with the Fermi surface evolution. 

%We also demonstrate that if nontopological $s$-wave pairing is realized in the bulk, the surface superconductivity is anomalously enhanced.
%by the parity mixing mechanism through the well-defined surface Dirac fermions. 
%Hence, the conventional $s$-wave pairing in Cu$_x$Bi$_2$Se$_3$ is accompanied by double coherence peaks in surface density of states (SDOS) but does not support a simple U-shaped SDOS.
%Using a fitting with the spectrum of a classical $s$-wave
%superconductor led to a nontopological
%$s$-wave pairing symmetry of Cu$_{x}$Bi$_{2}$Se$_{3}$, which is
%contrary to previous experiments.

Whereas there have been several theoretical works on 
STIs,~\cite{hao11,hsieh12,yamakage12}
the self-consistent analysis of the surface pairing potential has
been lacking.
We here present the first self-consistent calculation for STIs 
that takes account of the interplay between bulk superconductivity
and surface Dirac fermions in equal footing.
We show that, if an $s$-wave pairing is
realized in the bulk, the pair potential is enhanced near the surface with an
anomalous parity mixing of the pairing symmetry.
The enhanced pair potential opens a large gap for surface Dirac fermions,
and thus
the resultant SDOS hosts an
extra coherent peak at the induced gap, in addition to a conventional peak at
the bulk gap.
We illustrate that the enhancement and the parity mixing are mediated by 
surface Dirac fermions themselves. This is essentially different from the mechanism 
of a subdominant pair potential emergent in unconventional superconductors.~\cite{Matsumoto,Fogel,Greene,Tanuma}
We also would like to mention that this effect is also distinct from the parity mixing effect for bulk topological superconductivity.~\cite{TYBN09,satoPRB2009,TMYYS10,YSTY10,schnyderPRB2011,schnyderPRB2012}

We also examine surface states for bulk topological odd-parity pairing.
In contrast to bulk $s$-wave pairing,
neither mixture of a subdominant pair potential
nor gap opening of surface Dirac fermions occurs. It is demonstrated that 
the Fermi surface evolution from spheroidal to cylindrical shape induces 
a topological phase transition at which SABSs disappear. 
Based on our self-consist analysis, therefore, the bulk odd-parity pairing gives 
a consistent understanding 
on the ZBCP~\cite{koren11,kirzhner12,koren12}
and the U-shaped SDOS~\cite{peng13,Levy} observed in Cu$_x$Bi$_2$Se$_3$.

%%%%%%%%%%%%%%%%%%%%%%%%%%%%%%%%%%%%%%%%%%%%%%%%%%%%%%%%%%%%%%%%%%%%%%%%%%%%%%%%% Please prepare formulation  Figure and results
%
%-------------------- Nambu-Gor'kov formalism
\section{Self-consistent theory}

We start with the Hamiltonian for STIs, 
$\mathcal{H} \!=\! \int d{\bm r}\psi^{\dag}_{\alpha}({\bm r})[\mathcal{H}_{\rm TI}(-i{\bm \nabla}) ]_{\alpha\beta}\psi _{\beta}({\bm r}) + \int d{\bm r} [ U \{n^2_1({\bm r}) + n^2_2({\bm r})\} + 2V n_1 ({\bm r})n_2 ({\bm r})]$, proposed in Ref.~\onlinecite{fu10}, where $\psi _{\alpha}$ and $\psi^{\dag}_{\alpha}$ are fermionic field operators, and the repeated Greek indices imply the sum over the orbital $\sigma = 1, 2$ and spin $s = \uparrow,\downarrow$. The electron density operator is defined as $n_{\sigma} \!=\! \sum _s
\psi^{\dag}_{\sigma, s}\psi _{\sigma, s}$ and $U$ and $V$ are intra- and
inter-orbital density-density interactions, respectively. 
For the single-particle Hamiltonian ${\mathcal{H}}_{\rm TI}$, we consider the following $k\cdot p$ Hamiltonian describing the band structure of topological insulators near the $\Gamma$ point~\cite{zhang2009}:
\begin{align}
\mathcal{H}_{\rm TI}({\bm k}) = c({\bm k}) + m({\bm k})\sigma _x
+ v_zk_z\sigma _y + v\sigma _z({\bm k}\times{\bm s})_z,
\label{eq:HTI}
\end{align}
where $m ({\bm k}) \!=\! m_0 + m_1 k^2_z + m_2 (k^2_x+k^2_y)$ with $m_1m_2 \!>\!
0$ and $c({\bm k}) \!=\! -\mu +c_1k^2_z + c_2(k^2_x+k^2_y)$ with the chemical potential $\mu$. We introduce the Pauli
matrices in the spin and orbital spaces, $s_{\mu}$ and $\sigma
_{\mu}$. 
As shown in Sec.~\ref{sec:generic}, Eq.~\eqref{eq:HTI} is a generic form of the Hamiltonian preserving the mirror and $n$-fold rotation symmetries ($n\ge 2$) in addition to the inversion and time-reversal symmetries.
Therefore, Eq.~\eqref{eq:HTI} not only describes the band structure of Bi$_2$Se$_3$,~\cite{zhang2009} but also is applicable to other TIs, including SnTe.~\cite{fu}

The self-consistent electronic structure is obtained by solving 
the Bogoliubov-de Gennes equation
\beq
\underline{\mathcal{H}}(-i{\bm \nabla}) {\bm \varphi}_I({\bm r}) = E_I {\bm \varphi}_I({\bm r}),
\label{eq:bdg_final}
\eeq
where the eigenvector ${\bm \varphi}_I\!=\!(u_{I,\sigma,\uparrow},u_{I,\sigma,\downarrow},v_{I,\sigma,\uparrow},v_{I,\sigma,\downarrow})^{\rm T}$ satisfies the condition $\int d{\bm r}{\bm \varphi}^{\dag}_I({\bm
r}){\bm \varphi}_J ({\bm r}) = \delta _{IJ}$. The $8\times 8$ Hamiltonian density $\underline{\mathcal{H}}({\bm r})$ is given by
\beq
\underline{\mathcal{H}}(-i{\bm \nabla})  \equiv \left(
\begin{array}{cc}
\mathcal{H}_{\rm TI}(-i{\bm \nabla}) & -i\hat{\Delta} ({\bm r}) s_y \\
is_y \hat{\Delta}^{\dag} ({\bm r}) & -\mathcal{\mathcal{H}}^{\ast}_{\rm TI}(-i{\bm \nabla})
\end{array}
\right).
\eeq
The $4\!\times\!4$ matrix of the pair potential, $\hat{\Delta}$, is obtained from\begin{align}
i\left[\hat{\Delta} ({\bm r})s_y \right]_{\alpha\beta}  = \mathcal{V}_{\alpha\beta}
\sum _{E_{I}>0} & \left[
u_{I,\alpha}({\bm r}) v^{\ast}_{I,\beta}({\bm r}) f(E_{I}) \right. \nn \\
& \left. + v^{\ast}_{I,\alpha}({\bm r}) u_{I,\beta}({\bm r}) f(-E_{I})
\right],
\label{eq:gap_final}
\end{align}
where $\mathcal{V}_{\alpha\beta}$ is $U$ ($V$) for intra-orbital (inter-orbital) interactions and we set $\alpha = (\sigma,s)$. The set of self-consistent equations is derived in Sec.~\ref{sec:sc}
We self-consistently solve Eqs.~(\ref{eq:bdg_final}) and (\ref{eq:gap_final}) with the discrete
variable representation.~\cite{mizushimaPRA2008v2, takahashi}
Corresponding to (111) surfaces of Cu$_x$Bi$_2$Se$_3$, which are
naturally cleaved in the crystal of this material,
we place the boundary condition ${\bm \varphi}({\bm r}) = {\bm 0}$ at
$z = 0$ (bottom surface) and $L$ (top surface).
We assume homogeneity in the $x$-$y$ plane
parallel to the surface.

\begin{figure}[b]
\includegraphics[width=80mm]{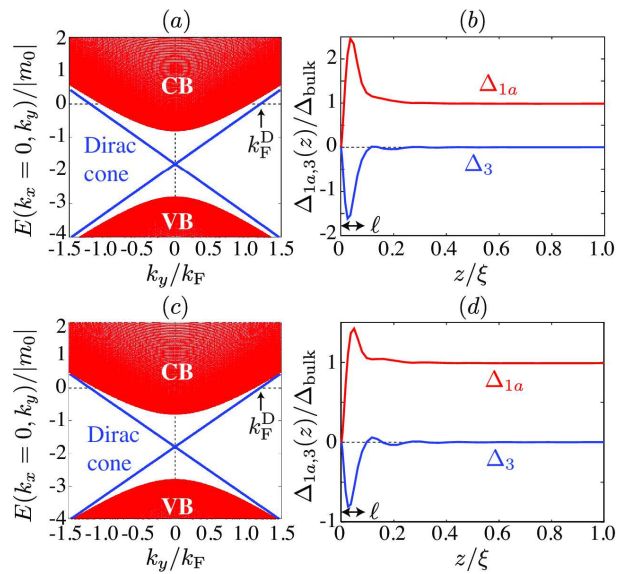}
\caption{(Color online) Energy spectra in the normal state with $\Delta = 0$ (a) and spatial profiles of $\Delta _{1a,3}(z)$ (b) for nontopological states for $\tilde{m}_2=-0.066$ ($\delta=0.159$) and $-0.20$ (0.08) (c, d). ``CB'' and ``VB'' denote the conduction and valence bands, respectively. We set $\tilde{m}_1 = -0.17$.}
\label{fig:gap}
\end{figure}

The material parameters are set as $m_0 \!=\! -0.28~{\rm eV}$, $v_z \!=\! 3.09 ~{\rm eV} \AA$, $v \!=\! 4.1 ~{\rm eV}\AA$, and $\mu/|m_0| \!=\! 1.8$.~\cite{sasaki11,yamakage12,LiuPRB10,zhang2009} For simplicity, we here focus on $c_1 \!=\! c_2\! =\! 0$ (The effect of nonzero $(c_1, c_2)$ is discussed in Sec.~\ref{sec:suppls}). We set $T \!=\! 0$, and $U$ and $V$ are chosen so as to fix the pair potential in the bulk, $\Delta _{\rm bulk}$. Although we deal with $\tilde{m}_1 \!\equiv\! m_1 m_0/v^2_z \!\in\! [-0.17,-0.59]$, $\tilde{m}_2\!\equiv\! m_2 m_0/v^2_z\in [-0.033,-0.33]$, and $\Delta _{\rm bulk}/|m_0| \!\in\! [0.01,0.1]$, we here focus on 
$\tilde{m}_1 \!= \! -0.17$ and $\Delta _{\rm bulk} \!=\! 0.1|m_0|$. To characterize the length scale of superconducting states, we define the coherence length as $\xi \!=\! v_{\rm F}/\Delta _{\rm bulk}$, which is estimated as $\xi \!=\! 12.5 k^{-1}_{\rm F}$. The Fermi momentum and velocity of the conduction band are determined by $E_{\rm CB}(k_{\rm F})=\mu$ and by $v_{\rm F} \!=\! \partial E_{\rm CB}/\partial k_{\parallel}|_{k_{\parallel}=k_{\rm F}}$, respectively, where the 
conduction band energy is $E_{\rm CB}\!=\! c_2 k^2_{\parallel}+\sqrt{(m^2_0+m_1k^2_{\parallel})^2 + v^2k^2_{\parallel}}$ 
with $k^2_{\parallel}\!=\!{k^2_x+k^2_y}$. 
For systematic study on the interplay between superconductivity and surface Dirac fermions, we introduce 
\beq
\delta \equiv \frac{k^{\rm D}_{\rm F}-k_{\rm F}}{k_{\rm F}},
\label{eq:separation}
\eeq 
which quantifies the separation between the conduction band and Dirac cone, where $k^{\rm D}_{\rm F}$ is the Fermi momentum of the Dirac cone (see Figs.~\ref{fig:gap}(a) and \ref{fig:gap}(c)). Since $k_{\rm F}$ is sensitive to $\tilde{m}_2$, the separation $\delta$ is controlled via $\tilde{m}_2$. The calculated range of $\tilde{m}_2$ corresponds to $0.05 \!<\! \delta \!<\! 0.18$, which is consistent with angle-resolved photoemission spectroscopy (ARPES) data.~\cite{wray}

%-------------------------------------------------
%------------------------------------------------/

%-------------------- pair potentials.

\section{Surface structure of superconducting topological insulators}

\subsection{Surface Dirac fermions of Cu$_x$Bi$_2$Se$_3$}

In Figs.~\ref{fig:gap}(a) and \ref{fig:gap}(c), we show the energy
spectra of Cu$_x$Bi$_2$Se$_3$ in the normal state.
Like the parent topological insulator Bi$_2$Se$_3$,
the topology of the normal state is characterized
by the $\mathbb{Z}_2$ invariant, $\nu$, which obeys 
$(-1)^{\nu} \!=\! {\rm sgn}(m_0m_1)$. For odd
$\nu$'s, topologically protected Dirac fermions are bound at the
surfaces. As seen in
Figs.~\ref{fig:gap}(a) and \ref{fig:gap}(c), at the Fermi level of
Cu$_{x}$Bi$_2$Se$_3$ (i.e., $E\!=\! 0$ in Fig.~\ref{fig:gap}), the Dirac cone is well isolated from the bulk
conduction band, which is consistent with ARPES data.~\cite{wray}
The separation $\delta$ increases with decreasing  magnitude of
$\tilde{m}_2$.
The wave function of Dirac fermions on the surface at $z\!=\! 0$ is solved for $c_1\!=\! c_2\!=\! 0$ as
\beq
{\bm \varphi}_{\rm D}(z) = (e^{-\kappa_- z}-e^{-\kappa_+ z})
\left(\begin{array}{c} 0 \\ 1 \end{array}\right)_{\sigma} \otimes u_s(k_x,k_y),
\label{eq:dirac}
\eeq
with the boundary condition ${\bm \varphi}_{\rm D}(0)\!=\!{\bm 0}$,
where
$\kappa_{\pm}\!=\!v_z/2m_1 \!\pm\!
\sqrt{(m_0+m_2k_{\parallel}^2)/m_1+(v_z/2m_1)^2}$
and
$(0,1)^{\rm T}_{\sigma}$ is the spinor in the orbital space. The spinor
$u_s$ in  spin space satisfies $(k_xs_y -
k_y s_x)u_s \!=\! s k_{\parallel} u_s$ with $s \!=\! \pm$.
${\bm \varphi}_{\rm D}$ is localized near the surface
for small $k_{\parallel}$.

Interestingly, the wave function of Eq.~(\ref{eq:dirac}) consists of
only one orbital, i.e., $\sigma=2$.
In other words,
surface Dirac fermions are fully polarized in the orbital space. 
This polarization is a consequence of the inversion symmetry breaking on
a surface:
In the bulk, the inversion symmetry,
$\mathcal{P}\mathcal{H}_{\rm TI}({\bm k})\mathcal{P}^{\dag} =
\mathcal{H}_{\rm TI}(-{\bm k})$ ($\mathcal{P} = \sigma _x$), ensures the
degeneracy in the orbital space, but on a surface, this
symmetry is completely broken so that  polarization can arise.
Although the $c_1$ and $c_2$ terms in Eq.~\eqref{eq:HTI} may weaken the orbital polarization of the Dirac cone,  it turns out that the effect of the $c_1$ and $c_2$ terms is negligible when $\delta$ is fixed.
The details are described in Sec.~\ref{sec:suppls}
As we will show below, the polarization has a significant influence on
 parity mixing and pairing symmetry near the surface.

%The orbital polarization also depends on the separation of the Dirac cone, $\delta \equiv (k^{\rm D}_{\rm F}-k_{\rm F})/k_{\rm F}$, where $k^{\rm D}_{\rm F}$ is the Fermi momentum of the Dirac cone.}

%----------------------------------------------------
\subsection{Surface states of bulk $s$-wave pairing}

We now turn to superconducting states. Because Fermi statistics
requires $\hat{\Delta}$ to satisfy $s_y\hat{\Delta}^{\rm T}s_y =
\hat{\Delta}$, there are six $k$-independent pairings in the bulk 
$(\Delta _{1a}$, $\Delta _{1b}\sigma_x$, $\Delta
_{2}\sigma_ys_z$, $\Delta_3\sigma_z$, $\Delta_{4x}\sigma_y s_x$, and
$\Delta_{4y}\sigma_y s_y)$,
whose symmetry properties are summarized in Table~I.
In the bulk, the short-range density-density interaction ${\cal H}_{\rm
int}$ realizes either $\Delta_{1a}+\Delta_{1b}\sigma_x$ or $\Delta_2\sigma_y
s_z$, depending on the model parameters $U$ and $V$.~\cite{fu10}
%Whereas the former belongs to the trivial representation $A_{1g}$
%of $D_{3d}$ corresponding to a topologically trivial $s$-wave
%pairing, the latter belongs
%to $A_{1u}$ realizing a fully gapped topological odd-parity
%pairing.
Whereas the former corresponds to $s$-wave pairing (the $A_{1g}$ representation for the $D_{3d}$ symmetry group of Cu$_x$Bi$_2$Se$_3$), the latter is a fully gapped odd-pairing (the $A_{1u}$ representation).

%-------------------------------------------------
\begin{table}
\begin{tabular}{c|c|c|c}
\hline\hline
Pair Potential & Inversion & Mirror & $\Gamma$ \\
\hline
$\Delta_{1a}$, $\Delta _{1b}\sigma _x$ & $+$ & $+$ & $A_{1g}$ \\
$\Delta_{2} \sigma_y s_z$  & $-$ & $-$ & $A_{1u}$ \\
$\Delta_{3}\sigma_z$  &  $-$ & $+$ & $A_{2u}$ \\
$\Delta_{4x,4y}\sigma_y s_{x,y}$ &  $-$ & $(+,-)$ & $E_u$\\
\hline\hline
\end{tabular}
\caption{Pairing potentials in STI and their parity under inversion $\mathcal{P}$ and mirror reflection $\mathcal{M}_{yz}$: $\Gamma$ denotes the representation of the $D_{3d}$ symmetry for Cu$_x$Bi$_2$Se$_3$.
}
\label{table1}
\end{table}
%------------------------------------------------/

Let us first examine the self-consistent surface state of the bulk $A_{1g}$
pairing, that is, a topologically trivial $s$-wave pairing $\Delta _{1a}$. 
For a while, we neglect the inter-orbital density-density interaction $V$ for simplicity.
%
%For a while, we neglect the inter-orbital density-density interaction $V$ and set
%$V=0$ for simplicity.
%In this case, the inter-orbital pairing $\Delta_{1b}\sigma_x$ is completely
%suppressed and only $\Delta_{1a}$ appears in the bulk.
%
Although the parity of the inversion is a good quantum number in the bulk,
it can be mixed near a surface since a surface breaks the inversion
symmetry.
Hence,  parity mixing of surface
pair potential may occur, though
the mixing pattern is restricted by symmetry surviving on the surface
considered.  
The mirror reflection with respect to the $y$-$z$ plane, ${\cal M}_{yz}=is_x$, restricts the
possible mixing as
\beq
\hat{\Delta}(z) = \Delta _{1a}(z) + \Delta _3(z)\sigma_z.
\label{eq:delta}
\eeq
The other pairings $\Delta_2$, $\Delta_{4x}$, and $\Delta_{4y}$ cannot
appear since they transform in a different manner than $\Delta_{1a}$
under the mirror reflection. 
Note that for $V\neq 0$, 
the possible mixing  of Eq.~(\ref{eq:delta}) is modified as
$\hat{\Delta} = \Delta_{1a} + \Delta _{1b}\sigma _x + \Delta _{3}\sigma _z$.
As seen in Fig.~\ref{fig:S4},
however, the induced $\Delta _{1b}$ does not alter our conclusion.
%the double-peak structure of the SDOS.

\begin{figure}[t]
\includegraphics[width=85mm]{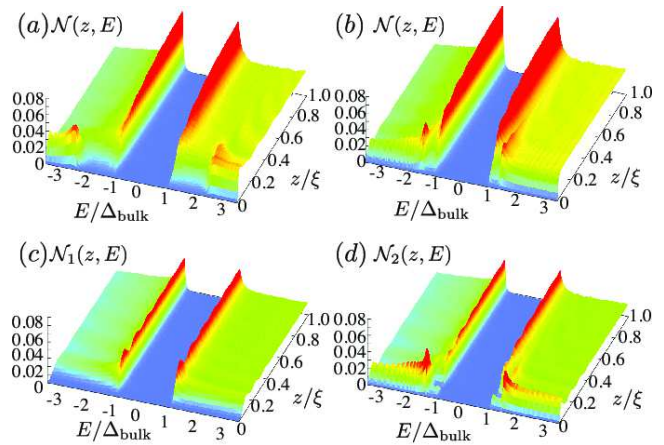}
\caption{(Color online) LDOS $\mathcal{N}(z,E)$ of the $A_{1g}$ state for $\tilde{m}_2=-0.066$ ($\delta = 0.159$) (a) and $-0.20$ ($0.08$) (b), where the pair potentials are shown in Figs.~\ref{fig:gap}(b) and \ref{fig:gap}(d). The LDOSs in $\sigma = 1$ and $\sigma = 2$ orbitals in the case of $\tilde{m}_2=-0.20$ are shown in (c) and (d).}
\label{fig:ldos}
\end{figure}

Figures~\ref{fig:gap}(b) and \ref{fig:gap}(d) show the self-consistently
obtained pair potential of Eq.~(\ref{eq:delta}), for
$\tilde{m}_2\!=\!-0.066$ ($\delta=0.159$) and $-0.20$ ($0.08$),
respectively.
Whereas the pair potential consists of only the $s$-wave
component $\Delta_{1a}$ in the bulk, the surface at $z=0$ induces
a large mixing of the odd-parity pairing $\Delta _3$.
Moreover, the $s$-wave pairing
$\Delta_{1a}$ itself
is strongly enhanced near the surface, being deviated from that of $\Delta
_{\rm bulk}$.
Both  the mixing and the enhancement are strengthened by decreasing
$|\tilde{m}_2|$, i.e., by increasing $\delta$, and they occur
near the surface within the scale of the penetration depth of the
Dirac cone, $\ell \!\equiv\! \kappa_-^{-1} \!\ll\! \xi$.

\begin{figure}[b]
\includegraphics[width=80mm]{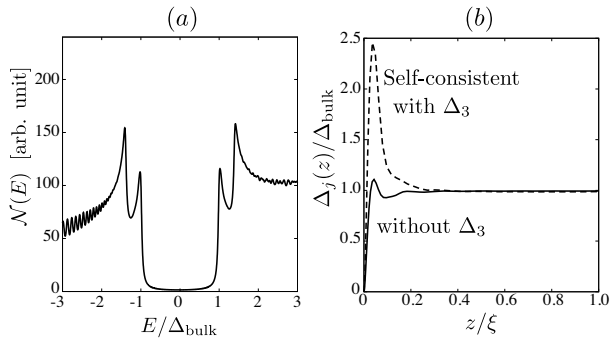}
\caption{(a) SDOS for the bulk $s$-wave pairing at $(\tilde{m}_1,\tilde{m}_2)=(-0.17,-0.20)$. (b) Comparison of spatial profiles of ${\Delta}_{1a}(z)$ with $V=0$ and $(\tilde{m}_1,\tilde{m}_2)=(-0.17,-0.066)$. In the solid curve, ${\Delta}_3$ is removed from the self-consistent iteration.}
\label{fig:delta3}
\end{figure}

The corresponding plots of local density of states (LDOS) $\mathcal{N}(z, E) \!\equiv\! \sum
_{\sigma}\mathcal{N}_{\sigma}(z,E)$
are shown in Figs.~\ref{fig:ldos}(a) and \ref{fig:ldos}(b), where
\begin{align}
\mathcal{N}_{\sigma}(z, E) = \sum_{I,s}[|u_{I,\sigma,s}|^2\delta (E-E_I)+|v_{I,\sigma,s}|^2\delta (E+E_I)].
\label{eq:ldos}
\end{align}
The LDOS is obtained from the analytic continuation of the Green's function
in Eq.~(\ref{eq:Gfinal}) as $
\mathcal{N}({\bm r},E) \!=\! -\frac{1}{\pi} {\rm Tr}_4G({\bm r},{\bm r};\omega _n \rightarrow -iE + 0_+)$, 
where $0_+$ is an infinitesimal constant. These figures clearly indicate the existence of an extra peak in the
SDOS at $E/\Delta _{\rm bulk} \approx \pm 2.5$ in Fig.~\ref{fig:ldos}(a)
[$E/\Delta_{\rm bulk}\approx\pm 1.5$ in Fig.~\ref{fig:ldos}(b)],
in addition to the bulk coherent peak at $E=\pm \Delta_{\rm bulk}$.
In Fig.~\ref{fig:delta3}(a), we display the averaged SDOS,
\beq
\mathcal{N}(E)\equiv \frac{1}{l_0}\int^{l_0}_0\mathcal{N}(z,E)dz,
\label{eq:sdos}
\eeq
where we set $l_0 = 2 k^{-1}_{\rm F}$.
The SDOS in the presence of surface Dirac fermions (Fig.~\ref{fig:delta3}(a)) 
sharply contrasts the SDOS of ordinary $s$-wave superconductors (see Fig.~\ref{fig:S5}(c)): 
Whereas the SDOS of ordinary $s$-wave superconductors supports only the
bulk coherent peak
and thus it is simply U-shaped, the present SDOS is not because of 
the enhancement of the surface superconductivity due to parity mixing.

Now we will argue a mechanism by which such anomalously large mixing and
enhancement of surface pair potential can occur in  STIs.
The key is the orbital polarization of surface Dirac fermions in Eq.~(\ref{eq:dirac}):
In Figs.~\ref{fig:ldos}(c) and \ref{fig:ldos}(d), we decompose the LDOS of
Fig.~\ref{fig:ldos}(b) into  two orbital components ${\cal N}_1$ and
${\cal N}_2$.
The decomposed LDOS clearly indicates that only the orbital $\sigma=2$
contributes to the enhanced pair potential near the surface.
This suggests that orbitally polarized surface Dirac fermions drive the
enhancement of the pair potential.
Note that the bulk quasiparticles cannot generate such a huge imbalance
between ${\cal N}_1$ and ${\cal N}_2$, since they are degenerate in the
orbital space.

Considering surface Dirac fermions, we can indeed explain
the qualitative behaviors of the surface pair potential in
Figs.~\ref{fig:gap}(b) and \ref{fig:gap}(d):
As was mentioned above,
surface Dirac fermions are fully polarized in the $\sigma \!=\! 2$ orbital
on the surface at $z\!=\!0$.
Therefore, near the surface, the $\sigma\!=\! 2$ component is dominated by
surface Dirac fermions, whereas the $\sigma\!=\! 1$ component consists of
only ordinary electrons from the bulk.
This means that these two components can behave differently near the
surface, and thus they determine the surface pair potential in different
manners.
In the $\sigma \!=\! 2$ component, a larger gap of surface Dirac fermions
is favored to gain the condensation energy of Dirac fermions,
but in the $\sigma \!=\! 1$ component, the surface pair potential should be
smoothly connected to the bulk pair potential.
As the $\sigma \!=\! 2$ ($\sigma \!=\! 1$) component of
Eq. (\ref{eq:delta}) is given by $\Delta_{1a}\!-\!\Delta_3$
($\Delta_{1a}\!+\!\Delta_3$), the former effect drives a nonzero
surface parity mixing term $\Delta_3$ opposite in sign to $\Delta_{1a}$. The latter, $\Delta_{1a}\!+\!\Delta_3$, must vanish at the surface, while it smoothly connects to $\Delta _{\rm bulk}$. 
This gives the condition of
$\Delta_{1a}\!+\!\Delta_3 \!\approx\! \Delta_{\rm bulk} \!>\! 0$.
The synergism between these two naturally leads to the enhancement
of $\Delta_{1a} \!\approx\! \Delta_{\rm bulk}\!+\!|\Delta_3|$ with a
large parity mixing $\Delta_3$.

To substantiate the above argument,
we demonstrate in Fig.~\ref{fig:delta3}(b)
how the pair potential behaves
if $\Delta _3$ is intentionally removed from the self-consistent
iteration.
This behavior clearly supports that  parity mixing is indispensable to the
enhancement of the surface pair potential.
Note also that the mixing and the enhancement of this mechanism
should  weaken as surface Dirac fermions are merged into the bulk, i.e., $\delta \! \rightarrow\! 0$.
This is also consistent with the difference between Figs.~\ref{fig:gap}(b)
and \ref{fig:gap}(d) as well as that between Figs.~\ref{fig:ldos}(a) and \ref{fig:ldos}(b).

%-------------------- Surface density of states.
\subsection{Surface states of bulk odd-parity pairing}

\begin{figure}[t]
\includegraphics[width=85mm]{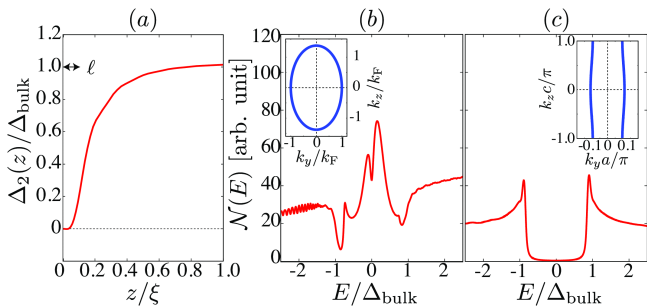}
\caption{(Color online) Spatial profiles of pair potential (a) and SDOS (b) for the $A_{1u}$ odd-parity state with the separation $\delta = 0.08$. (c) SDOS in  the $A_{1u}$ state with a cylindrical Fermi surface. The insets in (b) and (c) show the Fermi surface in the $k_y$-$k_z$ plane, where $a$ and $c$ are the lattice constants.}
\label{fig:topo}
\end{figure}

We also evaluate the self-consistent surface state for the
bulk $A_{1u}$  odd-parity superconductor.~\cite{footnote2}
As illustrated in Fig.~\ref{fig:topo}(a),
the self-consistently determined pair potential shows neither
 parity mixing nor  enhancement,
in contrast to the case of the $A_{1g}$ state (Fig.~1).
This is because being consistent with crystal symmetry on the surface
prohibits  parity mixing.
The spatial dependence of the pair potential is merely a typical one in the
presence of zero-energy SABSs.~\cite{Hara,Sauls,kashiwaya00}
The resulting SDOS in Fig.~\ref{fig:topo}(b)
is qualitatively the
same as that in non-self-consistent
calculations (see also Sec.~\ref{sec:odd}).~\cite{sasaki11, yamakage12}
Hence, the previous theoretical tunneling spectroscopy calculations,
which are consistent with  point-contact
experimental data having ZBCP,~\cite{sasaki11,koren11,kirzhner12,koren12}
are justified.

In Fig.~\ref{fig:topo}(c), we plot the SDOS in the $A_{1u}$ state with a cylindrical Fermi surface, calculated from the tight-binding model on the hexagonal lattice (see Sec.~\ref{sec:hexa}). This yields a simple U-shaped form. This is because the time-reversal invariant point $\Gamma$ is not enclosed by the Fermi surface on the (111) axis, so the Fermi surface evolution from the spheroidal to cylindrical shape induces the topological phase transition.~\cite{Sato09, Sato10} This feature is also observed in the $E_u$ state. 

%%%%%%%%%%%%%%%%%%%%%%%%%%%%%%%%%%%%%%%%%%%%%%%%%%%%%%%%%%%%%%%%%%%%%%%%%%%%%%%%

%-------------------- Conclusions.
\section{Concluding remarks}

In summary, we have studied the self-consistent structure of STIs.
We have found that,
if the bulk pairing symmetry is of $s$-wave type, surface Dirac fermions
induce a large magnitude of surface pair potential
accompanied by parity mixing.
As a result, the SDOS hosts an extra coherent peak at the induced gap as
well as that at the bulk gap.
Since the anomalous enhancement of surface superconductivity is a direct consequence of 
well-isolated surface Dirac fermions and parity mixing through the subdominant pair potential, 
our theory is applicable equally to other STIs.~\cite{sasakiPRL2012,sasakiPRB2013,arpino2013}
%although well-isolated surface Dirac fermions at the Fermi energy are
%necessary.

The present result has a direct implication on a recent STM experiment
on Cu$_x$Bi$_2$Se$_3$.~\cite{Levy}
Based on our theory, the classical U-shaped tunneling spectrum reported
in Ref.~\onlinecite{Levy} does not suggest an $s$-wave pairing of STI,
contrary to the previous claim.
Instead, we suggest here that the simple spectrum in Cu$_x$Bi$_2$Se$_3$ is
related to the evolution of the Fermi surface 
reported recently~\cite{dagan13,lawson}:
We have demonstrated that no gapless SABS appears on the (111) surface even
for the $A_{1u}$ odd-parity pairing, when the two-dimensional cylindrical Fermi surface is
realized in the bulk.~\cite{Sato09, Sato10}
At the same time, neither parity mixing nor enhancement of pair potential occurs for the
$A_{1u}$ pairing in both two- and three-dimensional Fermi surfaces.
Therefore, the two-dimensional odd-parity pairing naturally
reproduces the simple U-shaped spectrum of the STM experiment.

\begin{acknowledgments}

We thank Y. Ando for useful discussions.
This work is supported by JPSJ (Grants No.~25287085 and No.~25800199) and
the ``Topological Quantum Phenomena" Grant-in Aid (No.~22103005) for Scientific Research on Innovative Areas from MEXT of Japan.

%This work was supported by JSPS
%(No.~2074023303, 2134010303 and 22540383) and the MEXT KAKENHI (No.~22103002 and
%No.~22103005).
%``Topological Quantum Phenomena'' (No.~22103005) KAKENHI on innovation areas from MEXT.
% 21340103: Machida
% 25800199: Mizushima (Watate-B)

\end{acknowledgments} 

\appendix

\section{Self-consistent theory for superconducting topological insulators}
\label{sec:sc}

In this section, we describe the details on the derivation of the BdG equation (\ref{eq:bdg_final}) and gap equation (\ref{eq:gap_final}) for superconducting topological insulators. We also illustrate the possible pairing symmetry in Cu$_x$Bi$_2$Se$_3$ and a generic form of the Hamiltonian of carrier-doped topological insulators.

\subsection{Gor'kov equation}

We here explain the self-consistent equations which we have
used in this letter. 
We start with the following Hamiltonian for
spin-$1/2$ fermions with orbital degrees of freedom, 
\begin{align}
\mathcal{H} & =
\int d {\bm r} \psi^{\dag}_{\alpha}({\bm r}) \left[ H_{\rm TI}(-i{\bm \nabla})\right]_{\alpha\beta}\psi _{\beta}({\bm r}) \nn \\
& + 
\frac{1}{2}\int d{\bm r}_1 \int d{\bm r}_2 
\mathcal{V}^{\gamma,\delta}_{\alpha,\beta} (r_{12}) 
\psi^{\dag}_{\alpha}({\bm r}_1)\psi^{\dag}_{\beta}({\bm r}_2)
\psi _{\gamma}({\bm r}_2)\psi _{\delta}({\bm r}_1),
\label{eq:h0}
\end{align}
where $r_{12} = |{\bm r}_{12}| \!\equiv\! |{\bm r}_1-{\bm r}_2|$ is the relative coordinate and
$H_{\rm TI}$ describes the effective Hamiltonian for doped topological
insulators. The repeated Greek indices $\alpha$, $\beta$, $\delta$, and
$\gamma$ imply the sum over the orbital ($\sigma$) and spin $(s)$
degrees of freedom of electrons: $\alpha \!\equiv\! (\sigma,s)$.  
The Matsubara Green's functions are defined as $\underline{G}(x_1,x_2) \!=\! - \langle T_{\tau} [{\bm \Psi}(x_1){\bm \Psi}^{\dag}(x_2)] \rangle$ and
\begin{align}
\underline{G}(x_1,x_2) & = 
\left(
\begin{array}{cc}
\mathcal{G}(x_1,x_2) & \mathcal{F}(x_1,x_2) \\ \bar{\mathcal{F}}(x_1,x_2) & -\bar{\mathcal{G}}(x_1,x_2)
\end{array}
\right),
\end{align}
where $x_j \!\equiv\! (\tau_j, {\bm r}_j)$ and $\langle\cdots\rangle
\!\equiv\! {\rm Tr}[e^{(\Omega+\mu N-\mathcal{H})/T}\cdots]$ with the
thermodynamic potential $\Omega$. Here we have introduced the field
operator in Nambu space as
${\bm \Psi} = (\psi _{\sigma,\uparrow},\psi _{\sigma,\downarrow},\psi^{\dag}_{\sigma,\uparrow},\psi^{\dag}_{\sigma,\downarrow})^{\rm T}$.
Below, we expand $\underline{G}$ by the Matsubara frequency $\omega _n
\!=\! (2n + 1)\pi T$,
$\underline{G}(x_1,x_2) = T\sum _n \underline{G}({\bm r}_1,{\bm r}_2;\omega _n)e^{-i\omega _n \tau_{12}}
$. 

The Gor'kov equation is derived from the Heisenberg's equation of motion for fermionic field operators, $\frac{\partial}{\partial \tau}\psi _{\alpha}(x_j) = [\mathcal{H},\psi _{\alpha}(x_j)]$, as
\begin{align}
\int d{\bm r}_3
\left[
i\omega _n \delta ({\bm r}_{13}) - \underline{\mathcal{H}}({\bm r}_1,{\bm r}_3)
\right] \underline{G}({\bm r}_3,{\bm r}_2;\omega _n) = \delta ({\bm r}_{12}),
\label{eq:NG}
\end{align}
where $\underline{\mathcal{H}}$ is a $8\!\times\!8$ matrix in Nambu space,
\begin{align}
\underline{\mathcal{H}}({\bm r}_1,{\bm r}_2) = 
\left(
\begin{array}{cc}
\delta({\bm r}_{12})\mathcal{H}_{\rm TI}(-i{\bm \nabla}_1)  &
-i\hat{\Delta} ({\bm r}_1,{\bm r}_2) s_y \\
i\hat{\Delta}^{\ast}({\bm r}_1,{\bm r}_2) s_y & 
- \delta({\bm r}_{12})\mathcal{H}^{\ast}_{\rm TI}(-i{\bm \nabla}_1) 
\end{array}
\right).
\label{eq:Hbdg}
\end{align}
Here we have omitted the diagonal part of the self-energy matrix.
The pair potential is defined by the anomalous Green's functions,
\begin{align}
& \Delta _{\alpha\beta} ({\bm r}_1,{\bm r}_2)  
= - i\mathcal{V}^{\gamma\delta}_{\alpha\beta}({\bm r}_{12})
\left[\mathcal{F}({\bm r}_2,{\bm r}_1;\tau _{12}=0_+)s_y\right]_{\gamma\delta}
\nn \\
&=
- i\lim _{\eta \rightarrow 0}
T \sum _n \mathcal{V}^{\gamma\delta}_{\alpha\beta}({\bm r}_{12})
\left[\mathcal{F}({\bm r}_2,{\bm r}_1;\omega _n)s_y
\right]_{\gamma\delta}e^{-i\omega _n \eta}.
\label{eq:gap}
\end{align}
The Nambu-Gor'kov equation (\ref{eq:NG}) and the gap equation
(\ref{eq:gap}) determine 
the pair potential in a
self-consistent manner.

\subsection{Bogoliubov-de Gennes equation }

Now we show that the Gor'kov equation is reduced to the
Bogoliubov-de Gennes (BdG) equation (\ref{eq:bdg_final}),
\begin{eqnarray}
\int d{\bm r}_2 \underline{\mathcal{H}}({\bm r}_1,{\bm r}_2)
{\bm \varphi}_{I}({\bm r}_2)
= E_{I} {\bm \varphi}_{I}({\bm r}_1).
\label{eq:bdg3}
\end{eqnarray}
Here the eigenvector ${\bm \varphi}_{I}=(u_{I,\sigma,\uparrow},
u_{I,\sigma,\downarrow}, v_{I,\sigma,\uparrow},
v_{I,\sigma,\downarrow})^{\rm T}$  fulfills the orthonormal condition,
$\int {\bm \varphi}^{\dag}_{I}({\bm r}){\bm \varphi}_{J}({\bm r}) d{\bm
r} = \delta _{IJ}$.  
We first note that 
the BdG Hamiltonian Eq.~(\ref{eq:Hbdg}) is particle-hole symmetric,
$
\mathcal{C}\underline{\mathcal{H}}({\bm r}_1,{\bm r}_2)\mathcal{C}^{-1} = 
-\underline{\mathcal{H}}({\bm r}_2,{\bm r}_1)
$, where $\mathcal{C} = \underline{\tau}_x K$ with $K$ being the complex
conjugation operator. 
The  particle-hole symmetry of the BdG Hamiltonian ensures that the
positive energy solution ${\bm \varphi}_{E}({\bm r})$ is
associated with the negative energy solution ${\bm \varphi}_{-E}({\bm
r}) = \mathcal{C}{\bm \varphi}_{E}({\bm r})$.  
Therefore, the following $8\times 8$ unitary matrix
\begin{align}
\underline{u}_{I}({\bm r}) \equiv& [{\bm \varphi}^{(1)}_{I}({\bm
r}),{\bm \varphi}^{(2)}_{I}({\bm r}),{\bm \varphi}^{(3)}_{I}({\bm
r}),{\bm \varphi}^{(4)}_{I}({\bm r}), \nn \\
& \mathcal{C}{\bm
\varphi}^{(1)}_{I}({\bm r}), \mathcal{C}{\bm \varphi}^{(2)}_{I}({\bm
r}), \mathcal{C}{\bm \varphi}^{(3)}_{I}({\bm r}), \mathcal{C}{\bm
\varphi}^{(4)}_{I}({\bm r})] 
\end{align}
diagonalizes the BdG Hamiltonian as
$\int d{\bm r}_1 \int d{\bm r}_2\underline{u}^{\dag}_{I}({\bm r}_1) \underline{\mathcal{H}}({\bm r}_1,{\bm r}_2)
\underline{u}_{I}({\bm r}_2) = \underline{E}_{I}$, 
with
$\underline{E}_{I} \!\equiv\! {\rm diag}( 
E^{(1)}_{I}, E^{(2)}_{I},E^{(3)}_{I}, E^{(4)}_{I}, -E^{(1)}_{I}, -E^{(2)}_{I},
-E^{(3)}_{I}, -E^{(4)}_{I})$. 
The unitary matrix $\underline{u}_{I}({\bm r})$ satisfies the orthonormal
and completeness conditions,  
$\int \underline{u}^{\dag}_{I}({\bm r}_1) \underline{u}_{J}({\bm r}_1)
d{\bm r}_1 
= \delta _{IJ}\underline{\tau}_0$ and  
$\sum _{I} \underline{u}_{I}({\bm r}_1) \underline{u}^{\dag}_{I}({\bm r}_2)
= \delta ({\bm r}_{12})\underline{\tau}_0$.

By using the unitary matrix $\underline{u}_{I}$, the solution of the Gor'kov equation
(\ref{eq:NG}) is obtained as, 
$\underline{G}({\bm r}_1,{\bm r}_2; \omega _n) \!=\! \sum _{I} \underline{u}_{I}({\bm r}_1) 
\left( i\omega _n \underline{\tau}_0 - \underline{E}_{I} 
\right)^{-1} \underline{u}^{\dag}_{I}({\bm r}_2)
$, 
which can be recast into
%K
%The expression of $\underline{G}$ in the real coordinate is obtained from Eq.~(\ref{eq:G2}). 
%  ${\bm \varphi}_{I}$, $E_{I}$, it is recast into
\begin{align}
\underline{G}({\bm r}_1,{\bm r}_2; \omega _n)
= \sum _{E_I>0} \bigg[ &
\frac{{\bm \varphi}_{I}({\bm r}_1){\bm \varphi}^{\dag}_{I}({\bm r}_2)}{i\omega _n - E_{I}} \nn \\
& + 
\frac{\mathcal{C}{\bm \varphi}_{I}({\bm r}_1){\bm \varphi}^{\dag}_{I}({\bm r}_2)\mathcal{C}^{-1}}
{i\omega _n + E_{I}}
\bigg].
\label{eq:Gfinal}
\end{align}
From this expression,
the sum over the Matsubara frequency in Eq.(\ref{eq:gap}) results in the
Fermi distribution function $f(x) \!\equiv\! 1/(e^{x/T}+1)$, and thus,
the gap equation is reduced to
\begin{align}
i\left[\hat{\Delta} ({\bm r}_1,{\bm r}_2)s_y\right]_{\alpha\beta} 
&= \mathcal{V}^{\gamma\delta}_{\alpha\beta}({\bm r}_{12})
\sum _{E_{I}>0}  \left[
u_{I,\delta}({\bm r}_1) v^{\ast}_{I,\gamma}({\bm r}_2) \right. \nn \\
& \left. \times f(E_{I}) 
+ v^{\ast}_{I,\delta}({\bm r}_1) u_{I,\gamma}({\bm r}_2) f(-E_{I})
\right]. 
\label{eq:gap2}
\end{align}
We solve
the BdG equation Eq.(\ref{eq:bdg3}) and the gap equation
(\ref{eq:gap2}) self-consistently, instead of solving the Gor'kov
equation directly.

\subsection{Gap equations for Cu$_x$Bi$_2$Se$_3$}

In this paper, we consider the following short-range electron density-density
interaction as pairing interaction of superconducting topological
insulator Cu$_x$Bi$_2$Se$_3$~\cite{fu}:
\beq
\mathcal{H}_{\rm int} = U \left[ n^2_1 ({\bm r}) + n^2_2 ({\bm r})\right] + 2V n_1({\bm r})n_2 ({\bm r}),
\label{eq:Hint}
\eeq 
where the electron density operator in orbital $\sigma$ is defined as 
$n_{\sigma} = \sum _{s = \uparrow,\downarrow}
\psi^{\dag}_{\sigma,s}({\bm r})\psi _{\sigma,s}({\bm r})$, and $U$ and $V$
denote intra-orbital and inter-orbital interaction constant,
respectively.   
The general form of the pairing interaction $\mathcal{V}^{\gamma\delta}_{\alpha\beta}({\bm r}_{12})$ in Eq.~(\ref{eq:h0}) is simplified to  
$\mathcal{V}^{\gamma\delta}_{\alpha\beta}({\bm r}_{12}) =
\mathcal{V}_{\alpha\beta} \delta _{\alpha\delta} \delta _{\beta\gamma}
\delta ({\bm r}_{12})$, where 
the intra-orbital interaction ({\it i.e.}, $\sigma _{\alpha} = \sigma
_{\beta}$) yields $\mathcal{V}_{\alpha\beta} = U$ and
the inter-orbital one ($\sigma _{\alpha} \neq \sigma _{\beta}$) 
gives $\mathcal{V}_{\alpha\beta} = V$.
Using the effective pairing interaction, the gap equation (\ref{eq:gap_final}) 
is derived from Eq.~\eqref{eq:gap2} as
\begin{align}
i\left[\hat{\Delta} ({\bm r})s_y \right]_{\alpha\beta} 
=& \mathcal{V}_{\alpha\beta}
\sum _{E_{I}>0} \left[
u_{I,\alpha}({\bm r}) v^{\ast}_{I,\beta}({\bm r}) f(E_{I}) \right. \nn \\
& \left. 
+ v^{\ast}_{I,\alpha}({\bm r}) u_{I,\beta}({\bm r}) f(-E_{I})
\right].
\label{eq:gap_sti}
\end{align}
with $\Delta _{\alpha\beta} ({\bm r}_1,{\bm r}_2) = \Delta _{\alpha\beta}
({\bm r}_1) \delta ({\bm r}_{12})$. 
The BdG equation (\ref{eq:bdg3}) and gap equation (\ref{eq:gap_sti}) form a set of self-consistent equations for superconducting topological insulators.

The Fermi statistics of electrons imposes  the condition $\hat{\Delta} =
s_y \hat{\Delta}^{\rm T} s_y$, on the pair potential
$\hat{\Delta}$. 
There are six independent matrices,  
$(\Delta _{1a},\Delta _{1b}\sigma_x, \Delta _2\sigma_y
s_z, \Delta _3\sigma_z, \Delta _{4x}\sigma_ys_x, \Delta
_{4y}\sigma_ys_y)$, that satisfy this
condition.~\cite{fu,yamakage12} Hence, the general form of
$\Delta ({\bm r})$ is expanded in terms of the six independent pairings
\beq
\hat{\Delta} ({\bm r}) = \sum_{j} \Delta _j({\bm r}) \hat{\Gamma}_j.
\eeq 
Here the Hermitian matrices $\hat{\Gamma}_j \!=\! \hat{\Gamma}^{\dag}_j$ ($j
\!=\! 1a, 1b, 2, 3, 4x, 4y$) are given by, $\Gamma_{1a}=1_{4\times 4}$,
$\Gamma _{1b} \!=\!\sigma _x$, $\Gamma_2 \!=\! \sigma _y s_z$, $\Gamma_3 \!=\! \sigma
_z$, $\Gamma_{4x}\!=\!\sigma _y s_x$, and $\Gamma_{4y}\!=\!\sigma _y s_y$,
respectively. 
For the pair potential of Eq.~(\ref{eq:gap_sti}), 
the coefficients $\Delta _j$ are calculated by $\Delta _j
({\bm r}) \!=\! \frac{1}{4}{\rm Tr}_4[\hat{\Gamma}_j \hat{\Delta}({\bm r})]$.

In Table I, we summarize possible bulk pairing potentials of Cu$_x$Bi$_2$Se$_3$
and their properties.

\subsection{A generic form of the Hamiltonian preserving discrete symmetries}
\label{sec:generic}

We here derive a generic form of the Hamiltonian, $\mathcal{H}_{\rm TI}({\bm k})$, for spin-$1/2$ electron systems with two orbital degrees of freedom that preserves discrete symmetries, the inversion, time-reversal, mirror reflection, and $n$-fold rotation symmetries ($n\ge 2$). It is first convenient to introduce the $\gamma$ matrices that satisfy the following relations, 
$\{ \gamma _i, \gamma _j \} = 2\delta _{ij}$, where $i,j = 1, 2, \cdots, 5$. 
We also introduce their commutators $\gamma _{ij}$, which is defined as 
$\gamma _{ij} = \frac{1}{2i}[\gamma _i, \gamma _j] $. 
A generic form of a $4\times 4$ hermitian matrix can be expanded in terms of the five $\gamma$-matrices and ten their commutators in addition to the unit matrix as~\cite{murakamiScience2003,murakamiPRB2004}
\begin{align}
\mathcal{H}_{\rm TI}({\bm k}) = d_0({\bm k}) 
+ \sum _j d_j({\bm k})\gamma _j
+ \sum _{ij} d_{ij}({\bm k})\gamma _{ij}. 
\label{eq:h}
\end{align}

Let us now summarize the discrete symmetries relevant to topological insulators. Since the ${\mathbb Z}_2$ topological insulators hold the inversion symmetry and time-reversal symmetry,~\cite{fuPRB2007} the Hamiltonian must satisfy the following relations, 
\begin{gather}
\mathcal{P}\mathcal{H}_{\rm TI}({\bm k})\mathcal{P}^{\dag} 
= \mathcal{H}_{\rm TI}(-{\bm k}), \hspace{3mm} \mathcal{P} = \sigma _x, 
\label{eq:inv} \\
\mathcal{T}\mathcal{H}_{\rm TI}({\bm k})\mathcal{T}^{-1} 
= \mathcal{H}_{\rm TI}(-{\bm k}),  \hspace{3mm} \mathcal{T} = is_y K.
\label{eq:trs}
\end{gather}
We further suppose that the system holds the mirror reflection symmetry and $n$-fold rotation symmetry about the $\hat{\bm z}$-axis:
\begin{gather}
\mathcal{M}\mathcal{H}_{\rm TI}({\bm k})\mathcal{M}^{\dag} = \mathcal{H}_{\rm TI}(-k_x,k_y,k_z), \hspace{3mm}
\mathcal{M} = is_x,
\label{eq:mirror} \\
U_n \mathcal{H}_{\rm TI} ({\bm k}) U^{\dag}_n = \mathcal{H}_{\rm TI}(R_n{\bm k}),
\label{eq:rot}
\end{gather}
Here, the mirror reflection plane is set to be normal to the $\hat{\bm x}$-axis. A ${\rm SU}(2)$ rotation matrix in spin space is given as $U_n= \exp(-i\varphi s_z/2)$, which describes $n$-fold discrete rotation about the $\hat{\bm z}$-axis by an angle $\varphi = 2\pi /n$ ($n\in \mathbb{Z}$). The corresponding ${\rm SO}(3)$ rotation matrix is given by $R_n$. Equation \eqref{eq:rot} is the joint rotation of spin and momentum spaces. The mirror symmetry \eqref{eq:mirror} and $n$-fold rotation symmetry \eqref{eq:rot} are relevant to electron systems with crystalline symmetry.

Following Ref.~\onlinecite{fuPRB2007}, we choose the $\gamma$ matrix to be even under $\mathcal{P}\mathcal{T}$, $\mathcal{PT} \gamma _j \mathcal{T}^{-1}\mathcal{P}^{-1}=\gamma _j$, since the Hamiltonian in normal states is invariant the combination of the inversion symmetry and time-reversal symmetry, $\mathcal{PT} \mathcal{H}_{\rm TI}({\bm k}) \mathcal{T}^{-1}\mathcal{P}^{-1}= \mathcal{H}_{\rm TI}({\bm k})$. Then, the five $\gamma$ matrices are given as
$(\gamma _1, \gamma _2, \gamma _3, \gamma _4, \gamma _5) = 
(\sigma _x, \sigma _y, \sigma _z s_x, \sigma _z s_y, \sigma _z s_z)$.
Using this expression, one finds that all $\gamma _{ij}$ are odd under $\mathcal{PT}$, $\mathcal{PT} \gamma _{ij} \mathcal{T}^{-1}\mathcal{P}^{-1}=-\gamma _{ij}$. Hence, the inversion symmetry and time-reversal symmetry require that all $d_{ij}({\bm k})$ in Eq.~\eqref{eq:h} vanish. As a result, the $\mathcal{PT}$-invariant Hamiltonian is parametrized with the unit matrix and five $\gamma$ matrices as
\beq
\mathcal{H}_{\rm TI}({\bm k}) = d_0({\bm k}) 
+ \sum _j d_j({\bm k})\gamma _j
\label{eq:h2}
\eeq
The transformation of $\gamma$ matrices under $\mathcal{P}$, $\mathcal{T}$, and $\mathcal{M}$ is summarized in Table~\ref{table1}. This implies that $d_0({\bm k})$ and $d_1({\bm k})$ are even on ${\bm k}$ and $\mathcal{T}$ and otherwise odd, i.e., $d_0(-{\bm k}) = d_0({\bm k})$ and 
\beq
d_{j}(-{\bm k}) = \left\{ 
\begin{array}{ll}
d_j({\bm k}) & \mbox{for $j=1$} \\ 
\\
-d_j({\bm k}) & \mbox{for $j=2,3,4,5$}
\end{array}
\right.
\eeq
The time-reversal symmetry requires $d_j({\bm k})$ to be real. The mirror symmetry imposes constraint on the coefficients $d_{j}({\bm k})$ as
\beq
d_{j}(-k_x,k_y,k_z) = \left\{ 
\begin{array}{ll}
d_j({\bm k}) & \mbox{for $j=1, 2, 3$} \\ 
\\
-d_j({\bm k}) & \mbox{for $j=4,5$}
\end{array}
\right.
\label{eq:mirrord}
\eeq
and $d_0(-k_x,k_y,k_z) = d_0({\bm k})$.

%{\renewcommand\arraystretch{1.5}
\begin{table}
\begin{tabular}{c|ccccc}
& $\gamma _1$ & $\gamma _2$ & $\gamma _3$ & $\gamma _4$ & $\gamma _5$ \\
\hline
$\mathcal{P}$ & $+1$ & $-1$ & $-1$ & $-1$ & $-1$ \\
$\mathcal{T}$ & $+1$ & $-1$ & $-1$ & $-1$ & $-1$ \\
$\mathcal{M}$ & $+1$ & $+1$ & $+1$ & $-1$ & $-1$
\end{tabular}
\caption{Parity of the $\gamma$ matrices under the inversion operator $\mathcal{T}$, time-reversal operator $\mathcal{T}$, and mirror operator $\mathcal{M}$.
}
\label{table1}
\end{table}
%}

Let us first consider the $d_5({\bm k})\gamma _5$ term in $\mathcal{H}({\bm k})$. Since the time-reversal and inversion symmetries require $d_5({\bm k})$ to be odd on ${\bm k}$, it can be expanded in the lowest order on $k$ as
$d_5 ({\bm k}) = \alpha _{5x} k_x + \alpha _{5y} k_y + \alpha _{5z} k_z + \mathcal{O}(k^3)$,  
where $\alpha _{5x}$, $\alpha _{5y}$, and $\alpha _{5z}$ are arbitrary real coefficients. 
The mirror symmetry imposes constraint \eqref{eq:mirrord}, which leads to $\alpha _{5y} = \alpha _{5z} =0$. The $n$-fold rotation symmetry transforms the $d_5({\bm k})\gamma _5$ to
$U_n \left[ d_5({\bm k})\gamma _5\right] U^{\dag}_n 
= \alpha _{5x} \left(k_x\cos\varphi - k_y\sin\varphi  \right) \gamma _5$.
Hence, for $\mathcal{H}({\bm k})$ to be invariant under $\mathcal{M}$ and $n$-fold rotation, the $d_{5}({\bm k})\gamma _5$ term must vanish, 
\beq
d_5({\bm k}) = 0, \hspace{3mm} \mbox{when $n\ge 2$}.
\eeq
Similarly, since $\gamma _2$ is odd under $\mathcal{P}$ and $\mathcal{T}$ and even under $\mathcal{M}$, the coefficient $d_2({\bm k})$ is parameterized in the lowest order on $k$ as 
$d_2({\bm k}) = \alpha _{2y}k_y + \alpha _{2z} k_z$.
The $n$-fold rotation symmetry, however, requires $\alpha _{2y}=0$, i.e., 
\beq
d_2({\bm k}) = \alpha _2 k_z.
\eeq

Hence, a generic form of the Hamiltonian preserving discrete symmetries is given in the lowest order on $k$ as
\beq
\mathcal{H}({\bm k}) = d_0({\bm k}) + d_1 ({\bm k})\sigma _x 
+ \left[ d_3 ({\bm k}) s_x + d_4({\bm k}) s_y \right] \sigma _z.
\eeq
The coefficients $(d_0, d_1, d_3, d_4)$ must satisfy the following condition to preserve the $n$-fold rotation symmetry,
$d_0({\bm k}) \!=\! d_0(R_n{\bm k})$,  
$d_1({\bm k}) \!=\! d_1(R_n{\bm k})$, and
\begin{align}
e^{-i\varphi s_z}\left[d_3 ({\bm k})s_x + d_4({\bm k}) s_y \right] = d_3(R_n{\bm k}) s_x + d_4(R_n{\bm k})s_y.
\label{eq:d3d4}
\end{align}
Owing to the inversion, mirror, and time-reversal symmetries, $d_3({\bm k})$ and $d_4({\bm k})$ are parameterized as
$d_3({\bm k}) \!=\! \alpha _{3y}k_y +\alpha _{3z}k_z $, and 
$d_4 ({\bm k}) \!=\! \alpha _4 k_x$
in the lowest order on $k$. Then, the condition \eqref{eq:d3d4} is recast into
$e^{-i\varphi s_z} \left[\left( \alpha _{3y}k_y +\alpha _{3z}k_z\right) s_x
+ \alpha _4 k_x s_y \right] 
\!=\! \alpha _{3y}\left( k_x \sin\varphi + k_y \cos\varphi\right) s_x
\!+\! \alpha _{3z}k_z s_x 
\!+\! \alpha _4 \left( k_x \cos\varphi - k_y \sin\varphi \right) s_y$.
For $n\ge 2$, the condition is satisfied when 
\beq
\alpha _{3y} = - \alpha _4 , \hspace{3mm} \alpha _{3z} = 0.
\eeq

\begin{figure*}
\includegraphics[width=140mm]{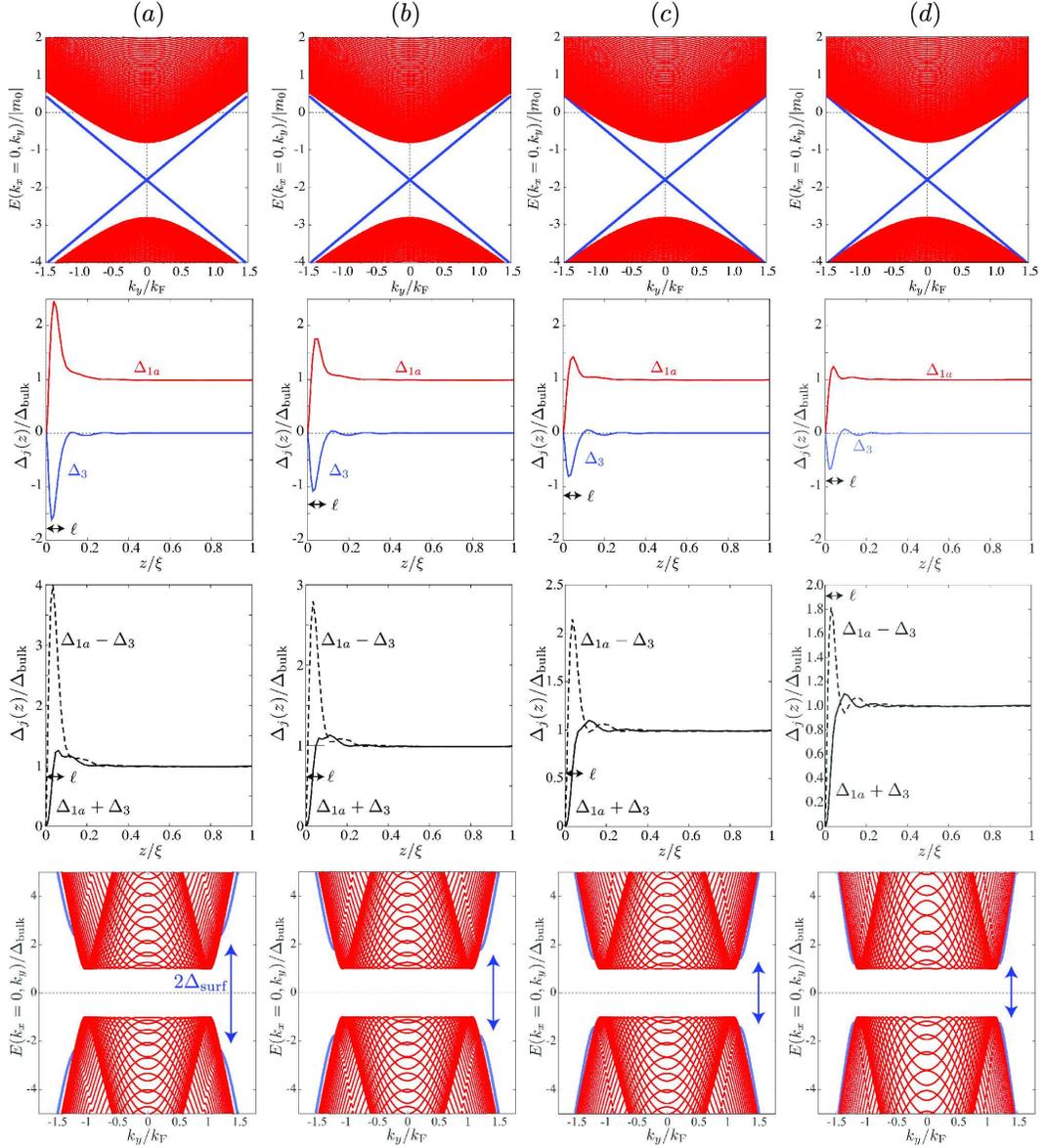}
\caption{(color online) The first row shows the energy spectra in the normal state for $\tilde{m}_1=-0.17$ and various $\tilde{m}_2$: $\tilde{m}_2 = -0.066$ ($\delta = 0.159$) (a), $-0.133$ ($\delta = 0.118$) (b), $-0.20$ ($\delta = 0.08$) (c), and $-0.266$ ($\delta = 0.04$) (d). In the second (third) row, we plot the spatial profiles of the pair potentials in the vicinity of the surface, $\Delta _{1a}$ and $\Delta _3$ ($\Delta _{1a} \pm \Delta _3$). Figures in the fourth row depicts  the quasiparticle energy spectra for the superconducting $A_{1g}$ state (bulk $s$-wave pairing state). Here we set $V=0$ and $\Delta _{\rm bulk} \equiv \Delta _{1a} (z=L/2) = 0.1 |m_0|$. The thick lines (blue color) show the dispersion originating from the surface Dirac fermions.} 
\label{fig:S1}
\end{figure*}

To this end, when a system preserves the inversion, time-reversal, mirror reflection, and $n$-fold rotation symmetries ($n\ge 2$), a generic form of the Hamiltonian is given in the lowest order on $k$ as
\begin{align}
\mathcal{H}_{\rm TI}({\bm k}) =& d_0({\bm k}) + d_1 ({\bm k})\sigma _x + \alpha _2 k_z \sigma _y \nn \\
& + \alpha _4 \left( k_x s_y - k_y s_x \right) \sigma _z,
\label{eq:hfinal}
\end{align}
where $d_0({\bm k})$ and $d_1 ({\bm k})$ which are even on $k$ are given by 
\begin{gather}
d_0 ({\bm k}) = \alpha _{00} + \alpha _{0\parallel}(k^2_x+k^2_y) + \alpha _{0\perp} k^2_z, \\
d_1 ({\bm k}) = \alpha _{10} + \alpha _{1\parallel}(k^2_x+k^2_y) + \alpha _{1\perp} k^2_z.
\end{gather}
Equation \eqref{eq:hfinal} has the same form as the $k\cdot p$ Hamiltonian describing the low-energy band structure of topological insulators, including Bi$_2$Se$_3$ and SnTe.~\cite{zhang2009,LiuPRB10,hsiehNP2012,fu} Hence, the form of the Hamiltonian (\ref{eq:HTI}) is generic for carrier-doped topological insulators with an appropriate crystalline symmetry.

\section{Supplementary numerical results}

In this section, we present supplementary numerical data of
self-consistently determined surface
pair potential.

\subsection{Surface states of bulk $s$-wave pairing}
\label{sec:suppls}

%\begin{figure}[b!]
%\includegraphics[width=70mm]{fig5.eps}
%\caption{(color online) Energy spectra in the normal state for $\tilde{m}_1=-0.17$ and various $\tilde{m}_2$: $\tilde{m}_2 = -0.066$ ($\delta = 0.159$) (a), $-0.133$ ($\delta = 0.118$) (b), $-0.20$ ($\delta = 0.08$) (c), and $-0.266$ ($\delta = 0.04$) (d). The thick lines (blue color) show the dispersion originating from the surface Dirac fermions.} 
%\label{fig:5}
%\end{figure}

First, the first row of Fig.~\ref{fig:S1} illustrates how the surface pair potential depends
on the separation of surface Dirac fermions from the bulk.
The separation becomes worse from left [(a)] to right
[(d)]. 
The first row of Fig.~\ref{fig:S1} shows the energy spectra in the normal 
state by solving $\mathcal{H}_{\rm TI}(-i{\bm \nabla}){\bm \varphi}_{\rm
D}({\bm r}) = E{\bm \varphi}_{\rm D}({\bm r})$, where the four-component
vector ${\bm \varphi}_{\rm D}({\bm r})$ obeys the boundary condition,
${\bm \varphi}_{\rm D}({\bm r}) = {\bm 0}$, on the surfaces $z=0$ and
$z=L$. We assume uniform infinite  $x$-$y$ plane. The wave function is
factorized as ${\bm \varphi}_{\rm D}({\bm r}) = e^{ik_xx+ik_yy}{\bm
\varphi}_{\rm D}(z)$.  
The effective Hamiltonian describing the band structure of Bi$_2$Se$_3$
near the $\Gamma$ point is given in Eq.~\eqref{eq:HTI},
where $m ({\bm k}) = m_0 + m_1 k^2_z + m_2 (k^2_x+k^2_y)$ with $m_1m_2 >
0$ and $c({\bm k}) = -\mu + c_1 k^2_z + c_2(k^2_x+k^2_y)$ with the chemical potential $\mu$.~\cite{zhang2009,LiuPRB10}
Here we set the parameters to be $m_0 = -0.28\; {\rm eV}$, $v_z = 3.09
\;{\rm eV} \AA$, $v=4.1 \; {\rm eV}\AA$, and $m_1 = 5.80~{\rm eV}\AA$
and the chemical potential is fixed to be $\mu/|m_0| =
1.8$.~\cite{sasaki11,yamakage12} The effect of the $c_1$ and $c_2$ terms are discussed in Fig.~\ref{fig:S5-1} and the corresponding text, and otherwise we set $c_1 \!=\! c_2 \!=\! 0$.
The parameter $m_2$ takes various values in the range of
$\tilde{m}_2 \!\equiv\! m_2 m_0/v^2_z \in[-0.033,-0.33]$. 
In the case of ${\rm sgn}(m_0m_1) <0$, the gapless Dirac cone exists, whose wave
function is bound on the surfaces within the penetration depth $\ell
\equiv \kappa^{-1}_-$ (see Eq.~\eqref{eq:dirac} and the subsequent sentences). To quantify the separation of the Dirac cone from the bulk conduction band, we introduce $\delta$ in Eq.~\eqref{eq:separation} that quantifies the separation of the surface Dirac cone from the conduction band. We find $0.05<\delta<0.18$ for the range of $-0.033<\tilde{m}_2<0.33$ for $\tilde{m}_1=-0.17$, which is consistent with angle-resolved photoemission spectroscopy data~\cite{wray}.
As illustrated in the first row of Fig.~\ref{fig:S1}, the surface Dirac
cone is
well isolated from the bulk spectrum at the Fermi level for small
$|\tilde{m}_2|$, i.e., large $\delta$.

In the second row of Fig.~\ref{fig:S1}, we show
corresponding numerical results of pair potentials near
the (111) surface in the superconducting state.
Here we have assumed $s$-wave pairing symmetry ($A_{1g}$ state of
Table~I) in the bulk.
The pair potentials are obtained by solving self-consistently
the BdG equation (\ref{eq:bdg_final}) and gap equation
(\ref{eq:gap_final}) in zero temperature.
We set 
$V=0$, and $U$ is chosen so as to fix
the pair potential in the bulk as $\Delta _{\rm bulk} = 0.1 |m_0|$, 
which corresponds to $k_{\rm F}\xi = 12.5$ in terms of the coherence length $\xi = v_{\rm F}/\Delta _{\rm bulk}$
with ``Fermi velocity'' $v_{\rm F} \!=\! \partial
E_{\rm CB}(k_{\parallel})/\partial k_{\parallel}|_{k_{\parallel}=k_{\rm F}}$.
The Fermi
momentum of the conduction band, $k_{\rm F}$, is determined by solving $E_{\rm CB}(k_{\rm F})=\mu$, where $k^2_{\parallel}\!=\!{k^2_x+k^2_y}$.
The results in the first row clearly indicate that the odd-parity  
pairing $\Delta _3$ is induced in the surface region within the length
scale of the penetration depth of the Dirac cone $\ell$. 

\begin{figure}[t!]
\includegraphics[width=70mm]{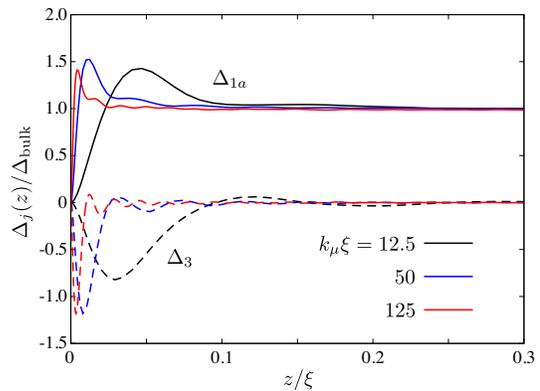}
\caption{(color online) Spatial profiles of the pair potentials in the non-topological $s$-wave 
pairing, $A_{1g}$ state, for various $k_{\mu}\xi$'s: $k_{\mu}\xi = 12.5$, $15$, and $125$. We here set $V=0$, $\tilde{m}_1=-0.17$, and $\tilde{m}_2=-0.20$ ($\delta = 0.08$). The solid and broken lines denote $\Delta _{1a} (z)$ and $\Delta _3 (z)$, respectively. The subdominant pair potential and the enhancement of $\Delta _{1a}$ on the surface are unchanged by increasing the dimensionless parameter $k_{\mu}\xi$, where the deviation and enhancement of the pair potentials are tightly bound at the length scale of the penetration depth of the surface Dirac fermion, $\ell \sim 2 k^{-1}_{\mu}$.}
\label{fig:S3}
\end{figure}

\begin{figure}[b!]
\includegraphics[width=80mm]{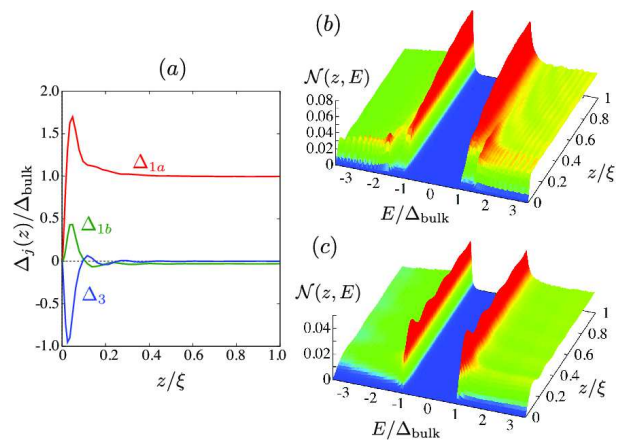}
\caption{Spatial profile of pair potentials (a) and LDOS (b) for the bulk $s$-wave superconducting $A_{1g}$ state with $V/U=1.0$ and $(\tilde{m}_1, \tilde{m}_2)=(-0.17,-0.20)$ ($\delta = 0.08$). (c) LDOS for the bulk $A_{1g}$ state with $\hat{\Delta} = \Delta _1$, where the underlying normal electrons are topologically trivial, $i.e.$, ${\rm sgn}(m_0m_1)=1$, which is not accompanied by the surface Dirac fermion.} 
\label{fig:S4}
\end{figure}

\begin{figure}[t!]
\includegraphics[width=70mm]{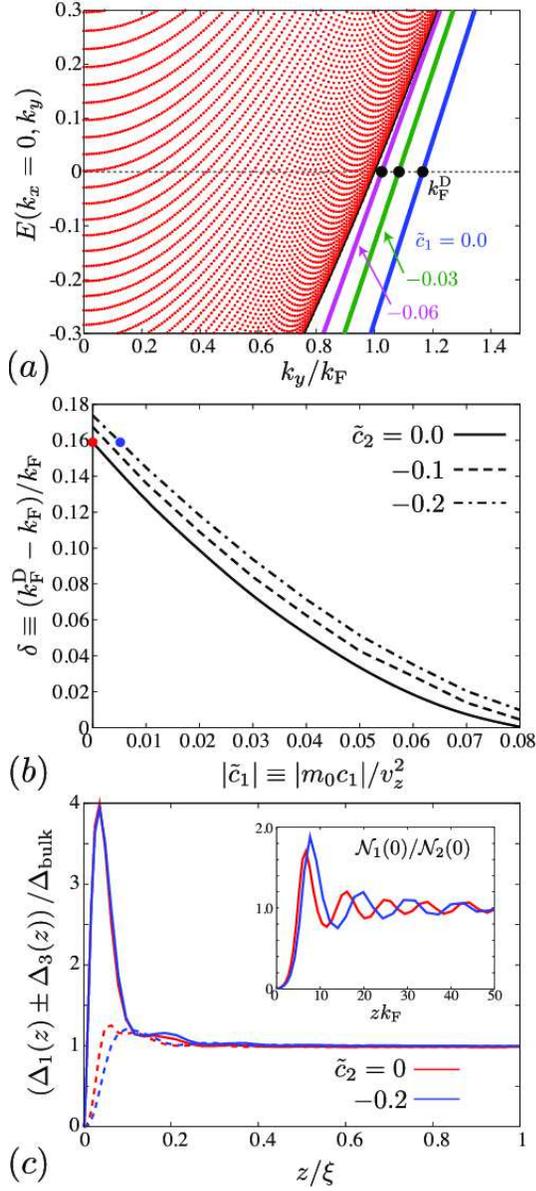}
\caption{(color online) (a) Energy spectra of the normal state for $\tilde{c}_1=0.0$, $-0.03$, and $-0.06$, where we set $\tilde{m}_1=-0.17$, $\tilde{m}_2=-0.066$, and $\tilde{c}_2=-0.1$ are fixed. (b) Difference $\delta$ between the Fermi surfaces of the conduction band $k_{\rm F}$ and the Dirac cone $k^{\rm D}_{\rm F}$ as a function of $\tilde{c}_1$ for $\tilde{c}_2 = 0.0$, $-0.1$, and $-0.2$.
The inset in (c) shows $\mathcal{N}_1(0)/\mathcal{N}_2(0)$ for $(\tilde{c}_1,\tilde{c}_2)=(0,0)$ and $(-0.0052,-0.2)$, where $\mathcal{N}_{\sigma}(0)$ is the zero-energy LDOS of the $\sigma$ orbital in the normal state.}
\label{fig:S5-1}
\end{figure}

The third row of Fig.~\ref{fig:S1} decomposes the same pair
potentials in the orbital components, $\Delta
_{1a}(z) \pm \Delta _3(z)$. 
It is seen that the pair potential $\Delta _{1a}-\Delta _3$ for the
$\sigma =2$ orbital is strongly enhanced near 
the surface, while the pair potential $\Delta _{1a}+\Delta _3$ for the
$\sigma =1$ orbital is not. 
This reflects that the surface Dirac fermions occupy only
one orbital state, $\sigma =2$ (see Eq.~\eqref{eq:dirac}). 

The quasiparticle energy spectra in the bulk
$s$-wave pairing state, $E(k_x,k_y)$, are displayed in the fourth row of
Fig.~\ref{fig:S1}.  
The quasiparticle spectra have a two-gap
structure, where a larger gap, $E = \pm \Delta _{\rm surf}$, opens for
the surface Dirac fermions in addition to the bulk gap $E = \pm \Delta _{\rm
bulk}$. The magnitude of the surface gap is enhanced with decreasing the
amplitude of $\tilde{m}_2$. The corresponding LDOS is displayed in Fig.~\ref{fig:ldos}.

%In Figs.~\ref{fig:S2}(a-d), we depict the local density of states (LDOS)
%for the self-consistent solutions displayed in Fig.~\ref{fig:6}. 
%The LDOSs in the vicinity of
%the surface indicate that the surface Dirac fermions support an energy gap,
%$\Delta _{\rm surf}$, larger than $\Delta _{\rm bulk}$. The surface gap
%becomes larger with decreasing $|\tilde{m}_2|$. 

We have numerically checked how the surface pair potential behaves for
weaker pairing $k_{\rm F}\xi\gg 1$. 
Figure~\ref{fig:S3} shows 
the self-consistently
obtained pair potentials 
for $k_{\rm F}\xi = 12.5$, 50, and 125.
% where
%the limit of $k_{\rm F}\xi \gg 1$ corresponds to the weak coupling
%limit. 
It turns out that the parity mixing and the enhancement of the
surface pair potential are robust for weak pairing.

Figures~\ref{fig:S4}(a)
and \ref{fig:S4}(b) illustrate the pair potentials and the LDOS for the
bulk $s$-wave pairing state with a finite inter-orbital coupling
$V/U=1.0$. The inter-orbital spin-singlet pairing $\Delta _{1b}$ is
induced by nonzero $V$ in the vicinity of the surface. Nevertheless, the
SDOS does not alter the double-peak structure, as illustrated in
Fig.~\ref{fig:S4}(b).

For comparison, we have also studied the case without surface Dirac
fermions. This case is realized by choosing a topologically trivial
normal state with ${\rm sgn}(m_0m_1)=+1$. The resultant LDOS is shown in
Fig.~\ref{fig:S4}(c), which yields a merely simple U-shape even in the vicinity
of the surface. This result supports our claim that the existence of
surface Dirac fermions is indispensable to the large parity mixing and
enhancement of the surface pair potential.

In Fig.~\ref{fig:S5-1}, we discuss the effect of the diagonal self-energy $c_1k^2_z + c_2k^2_{\parallel}$ in the $A_{1g}$ state. The term coupled to $c_2$ changes the Fermi radius $k_{\rm F}$ of the conduction band at $k_z=0$, while the $c_1$ term relatively shifts the Fermi momentum of the Dirac cone, as shown in Fig.~\ref{fig:S5-1}(a). Note that for Bi$_2$Se$_3$, the values of $\tilde{c}_1$ and $\tilde{c}_2$ are estimated as $-0.3$ and $-1.6$.~\cite{LiuPRB10} However, the Dirac cone is ill-defined at the Fermi level if we use the same parameters given in Ref.~\onlinecite{LiuPRB10}. The well-defined surface Dirac cone requires $|\tilde{c}_1|$ and $|\tilde{c}_2|$ to be sufficiently small in addition to the small $\tilde{m}_2$. 

\begin{figure}[t!]
\includegraphics[width=85mm]{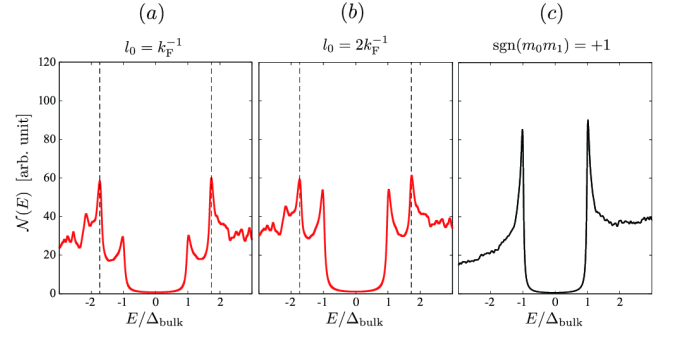}
\caption{(color online) SDOS, $\mathcal{N}(E)$, for the bulk $s$-wave superconducting states, where we set $k_{\rm F}\xi = 125$, $(\tilde{m}_1,\tilde{m}_2) = (-0.17,-0.20)$, and $V=0$ ($\delta = 0.08$). The definition for $\mathcal{N}(E)$ is given in Eq.~(\ref{eq:sdos}), where $l_0$ is set to be $l_0 = k^{-1}_{\rm F}$ (a) and $2k^{-1}_{\rm F}$ (b). (c) SDOS, $\mathcal{N}(E)$, for the bulk $s$-wave state without surface Dirac fermions, ${\rm sgn}(m_0m_1)=+1$, where we set $l_0 = 2k^{-1}_{\rm F}$.}
\label{fig:S5}
\end{figure}

The separation $\delta$ defined in Eq.~\eqref{eq:separation} is plotted in Fig.~\ref{fig:S5-1}(b) as a function of $\tilde{c}_1\!\equiv\! m_0 c_1/v^2_z$ for various $\tilde{c}_2$. It is seen that as $|\tilde{c}_1|$ increases, the surface Dirac cone is merged to the bulk conduction band. To clarify the effect of the $\tilde{c}_1$ and $\tilde{c}_2$ terms on the enhancement of the pair potential at the surface, we plot the spatial profiles of the $\Delta _{1a} \pm \Delta _{3}$ for $(\tilde{c}_1, \tilde{c}_2) = (0,0)$ and $(0.0052,-0.2)$, where the separation is fixed to be $\delta =0.159$. As seen in Fig.~\ref{fig:S5-1}, the ratio of the suface density of states (SDOS) in each orbital is slightly deviated by increasing $|c_1|$ and $|c_2|$, which indicates that the relative population of the orbitals in the surface Dirac cone varies. The spatial profile of the pair potentials at the surface is, however, insensitive to the increase of $|c_1|$ and $|c_2|$.

Finally, using Eq.~(\ref{eq:ldos}), we calculate the SDOS defined in Eq.~\eqref{eq:sdos}, 
which is a direct observable in STM experiments. The length scale $l_0$
denotes the probing depth and is of the atomic order $\sim
k^{-1}_{\rm F}$. In Fig.~\ref{fig:S5}, we plot $\mathcal{N}(E)$ for the
bulk $s$-wave superconducting state with $k_{\rm F}\xi = 125$, which
clearly indicates an
extra coherent peak at the surface gap $E= \pm \Delta _{\rm surf}\approx
1.6 \Delta _{\rm bulk}$, besides the conventional peak at the bulk
gap. 
In contrast, such a double-peak structure is never seen in the case without
the surface Dirac fermions (see Fig.~\ref{fig:S5}(c)), where the electron state in the normal state is a
topologically trivial, ${\rm sgn}(m_0m_1) = +1$.

\subsection{Bulk topological odd-parity pairing}
\label{sec:odd}

\begin{figure}[t!]
\includegraphics[width=85mm]{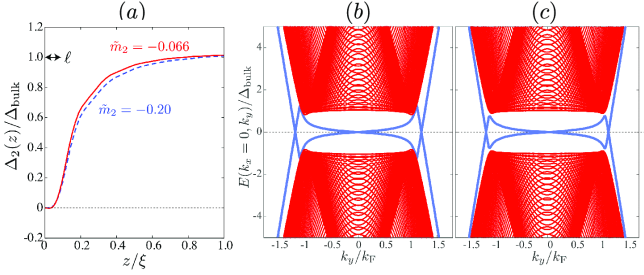}
\caption{(a) Spatial profiles of the pair potentials in the topological odd-parity state, the $A_{1u}$ state. The pair potential $\Delta _{\rm bulk}$ is determined by the amplitude at the center of the system, $\Delta _{\rm bulk}\equiv\Delta _{2} (z = L/2)$. The corresponding energy spectra of quasiparticles at $\tilde{m}_2 = -0.066$ ($\delta = 0.159$) and $-0.20$ ($\delta = 0.08$) are shown in (b) and (c), respectively.} 
\label{fig:S6}
\end{figure}

\begin{figure}[t!]
\begin{center}
\includegraphics[width=85mm]{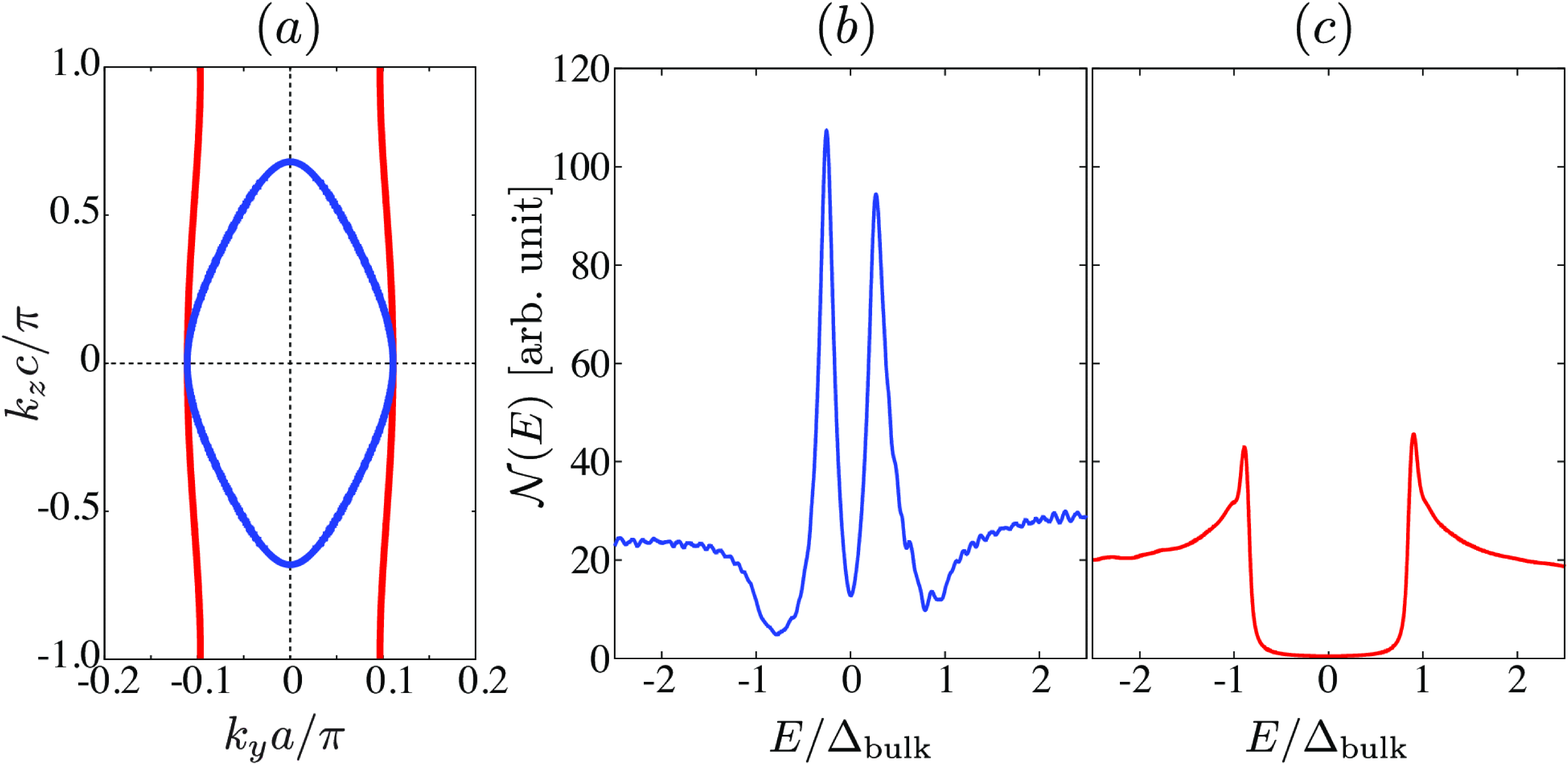}
\caption{(a) Shape of the Fermi surface: the spheroidal (the blue curve) and cylindrical (red) Fermi surfaces are given by changing the set of the parameters $(c_1, m_1, v_z)$. SDOS's in the $A_{1u}$ state with the spheroidal Fermi surface (b) and cylindrical Fermi surface (c).} 
\label{fig:S8}
\end{center}
\end{figure}

We now turn to the results for bulk topological odd parity superconductor, $\hat{\Delta}=\Delta _{2}\sigma _y s_z$.
As shown in Fig.~\ref{fig:S6}(a),
the self-consistently determined pair potential is suppressed near the surface region and the subdominant component never mixes, in contrast  to the case of the non-topological $s$-wave pairing. 
As shown in Figs.~\ref{fig:S6}(b) and \ref{fig:S6}(c), the resultant energy dispersions are qualitatively same as previous results obtained with a spatially uniform pair potential.~\cite{yamakage12,sasaki11,hsieh12,hao11}

\subsection{Tight-binding Hamiltonian and Fermi surface evolution}
\label{sec:hexa}

To clarify the effect of the Fermi surface evolution from the spheroidal to cylindrical shape, we introduce the tight-binding model for superconducting topological insulators. We consider a hexagonal lattice whose primitive vectors are $(\sqrt{3}a/2,a/2,0)$, $(0,a,0)$, and $(0,0,c)$. The tight-binding Hamiltonian is obtained from Eq.~\eqref{eq:HTI} by replacing ${\bm k}$ as follows~\cite{Hashimoto2013}:
$k_x \!\rightarrow\! \frac{2}{\sqrt{3}a} \sin\frac{\sqrt{3}k_x a}{2}\cos\frac{k_ya}{2}$, 
$k_y \!\rightarrow\! \frac{2}{3a}( 
\cos\frac{\sqrt{3}k_xa}{2} \sin\frac{k_ya}{2}
+ \sin k_y a
)$,  
$k^2_x + k^2_y \!\rightarrow\! \frac{4}{3a^2}
(
3 - 2\cos\frac{\sqrt{3}k_x a}{2}\cos\frac{k_ya}{2}-\cos k_ya
)$, 
$k_z \!\rightarrow\! \frac{1}{c} \sin k_z c$, and  
$k^2_z \!\rightarrow\! \frac{2}{c^2}(1-\cos k_zc)$,
where $a$ and $c$ are the lattice constants and for Bi$_2$Se$_3$, $a=4.14 \AA$ and $c = 28.7 \AA$. Using the tight-binding Hamiltonian, we self-consistently solve the BdG and gap equations in the $A_{1u}$ state which is the topological odd-parity pairing.

In Fig.~\ref{fig:S8}(a), we show the shape of the Fermi surface in the normal state. For the calculation of the SDOS presented in the main text, we set the parameters as follows: $m_0 =-0.28~{\rm eV}$, $\mu = 1.8 |m_0|$, $c_2 = 30.4~{\rm eV}\AA$, $m_2 = 44.5~{\rm eV}\AA^2$, and $v = 3.33 {\rm eV}\AA$ as given in Ref.~\onlinecite{LiuPRB10}. To change the shape of the Fermi surface, we choose $c_1/c^2 = 0.024~{\rm eV}$, $m_1/c^2=0.20~{\rm eV}$, and $v_z/c ~ 0.32~{\rm eV}$ for the spheroidal Fermi surface,~\cite{Hashimoto2013} and  $c_1/c^2 = 0.01~{\rm eV}$, $m_1/c^2=0.05~{\rm eV}$, and $v_z/c ~ 0.05~{\rm eV}$ for the cylindrical shape. The corresponding SDOS profile for the spheroidal Fermi surface is displayed in Fig.~\ref{fig:S8}(b). Here, the double low-energy peaks are observed, because these values of the parameters indicate that the surface Majorana cone is not twisted.~\cite{yamakage12} For the spheroidal Fermi surface, as shown in Fig.~\ref{fig:S8}(c), the surface state in the $A_{1u}$ state vanishes, resulting in the simple U-shaped LDOS on the surface.

\bibliography{proximity5}

\end{document}